\documentclass{llncs}
\usepackage{graphics,latexsym,amssymb}
\title{Mechanical generation of networks with surplus complexity}

\newcommand{\integers}{{\mathbb Z}}

\def\citeyear(#1)#2{\cite{#2}}

\newcommand{\EcoLab}{{\sffamily\slshape
    \mbox{\raisebox{.5ex}{Eco}\hspace{-.4em}{\makebox[.5em]{L}ab}}}}

\begin{document} 
\author{Russell K. Standish}
\institute{Mathematics and Statistics, University of New South Wales
                       \email{hpcoder@hpcoders.com.au}}

\maketitle 
\begin{abstract} 
In previous work I examined an information based complexity measure
of networks with weighted links. The measure was compared with that
obtained from by randomly shuffling the original network, forming an
Erd\"os-R\'enyi random network preserving the original link weight
distribution. It was found that real world networks almost invariably
had higher complexity than their shuffled counterparts, whereas
networks mechanically generated via preferential attachment did
not. The same experiment was performed on foodwebs generated by an
artificial life system, Tierra, and a couple of evolutionary ecology
systems, EcoLab and WebWorld. These latter systems often exhibited
the same complexity excess shown by real world networks, suggesting
that the {\em complexity surplus} indicates the presence of
evolutionary dynamics.

In this paper, I report on a mechanical network generation system
that does produce this complexity surplus. The heart of the idea is
to construct the network of state transitions of a chaotic dynamical
system, such as the Lorenz equation. This indicates that complexity
surplus is a more fundamental trait than that of being an evolutionary system.
\end{abstract}

\section{Introduction}

This work situates itself firmly within the {\em complexity is
information content} paradigm, a topic that dates back to the 1960s
with the work of Kolmogorov, Chaitin and Solomonoff. Indeed, the
seminal work of Mowshowitz in 1968
\cite{Mowshowitz68a,Mowshowitz68b,Mowshowitz68c,Mowshowitz68d}
describes an information-based network complexity measure called {\em
graph entropy}, that is essentially a generalisation of the measure
presented here, work that has by and large been forgotten by the
complex systems community, only to be reinvented in recent
times\cite{Dehmer-Mowshowitz14}.

The idea is fairly simple.  In most cases, there is an obvious {\em
prefix-free} representation language within which descriptions of the
objects of interest can be encoded.  There is also a classifier of
descriptions that can determine if two descriptions correspond to the
same object. This classifier is commonly called the {\em observer},
denoted $O(x)$.

To compute the complexity of some object $x$, count the number of
equivalent descriptions $\omega(\ell,x)$ of length $\ell$ that map to
the same object $x$ under the agreed classifier. Then the complexity of $x$
is given in the limit as $\ell\rightarrow\infty$:
\begin{equation}\label{complexity}
{\cal C}(x) = -\log P(x) = \lim_{\ell\rightarrow\infty} (\ell\log N - \log\omega(\ell,x))
\end{equation}
where $N$ is the size of the alphabet used for the representation
language. Loosely speaking, $P(x)$ here is the probability that a description
chosen uniformly at random will describe the object $x$.

To fix the representation language of graph using a binary alphabet (N=2), we start with a prefix of
$\lceil \log_2n \rceil$ 1s followed by a `0' stop bit. This indicates the
number of bits needed to store $n$, the number of nodes. Next we
encode the number of links $l$, which for a directed graph requires
$\lceil \log_2 n + \log_2(n-1) \rceil$ bits. Finally, we encode the
linklist, as a {\em rank index} within the 
\begin{displaymath}
\Omega=\left(\begin{array}{c} 
L \\
l \\
\end{array}
\right)
\end{displaymath}
possible linklists with $l$ links, where $L=n(n-1)$. So each
description of an $n$-node, $l$-link graph is precisely $\ell_{n,l}=1+2\lceil
\log_2n \rceil+\lceil \log_2 n + \log_2(n-1) \rceil + \lceil \log_2
\Omega \rceil$ bits long. Descriptions longer than that represent the
same graph as the initial leadin sequence just described, with the
trailing bits irrelevant, thus $\omega(\ell,x) =
\omega(\ell_{n,l},x)2^{\ell-\ell_{n,l}}$. Substituting into equation
(\ref{complexity}) gives (in bits):
\begin{equation}\label{graph-complexity}
{\cal C}(x) = \ell_{n,l} - \log_2\omega(\ell_{n,l},x).
\end{equation}


The relationship of this algorithmic complexity measure to more
familiar measures such as Kolmogorov (KCS) complexity, is given by the
coding theorem\cite[Thm 4.3.3]{Li-Vitanyi97}. In this case, the
descriptions are halting programs of some given universal Turing
machine $U$, which is also the classifier. Equation (\ref{complexity})
then corresponds to the logarithm of the {\em universal a priori
probability}, which is a kind of formalised Occam's razor that gives
higher weight to simpler (in the KCS sense) computable theories for
generating priors in Bayesian reasoning. The difference between this
version of ${\cal C}$ and KCS complexity is bounded by a constant
independent of the complexity of $x$, so these measures become
equivalent in the limit as message size goes to infinity.

In setting the classifier function, we assume that only the graph's
topology counts --- positions, and labels of nodes and links are not
considered important. Links may be directed or undirected and can have
a positive real number weight attached to them.

If all links have the same weight, then the counting problem of
determining $\omega(\ell_{n,l},x)$ for networks turns out to be equivalent
to the automorphism group problem of determining if two graphs are
automorphic. Whilst this problem is suspected of being combinatorially
hard\cite{Lubiw81}, several practical algorithms are available for
computing the size of the automorphism group for reasonable sized
networks of thousands of nodes.

To handle weighted links, the idea is to interpolate between the graph
with the link, and without the link, according to the link's
weight. The algorithm performs a weighted sum of graph complexity over
the different weights $w$ present in the network, of graphs composed
of links of weight greater than or equal to $w$. Details can be found
in \cite{Standish12}.

In \cite{Standish12}, this measure is applied to a number of
well-known naturally occurring networks, mostly published food
webs. The measure is also compared with an ensemble of networks
obtained by shuffling the link structure, with a positive complexity
difference ({\em complexity surplus}) usually found, indicating it is
measuring something structurally important about the network. The same
technique was applied to Ed\"os-R\'enyi generated networks, which
unsurprisingly had no complexity surplus, and to networks generated
via preferential attachment, which also generates no significant complexity
surplus.

In \cite{Standish10b}, I made the observation that artificial
evolutionary processes, such as EcoLab and Tierra also did not lead to
a complexity surplus, making this property of natural networks a
mysterious one. However, this later turned out to be due to a bug in
the analysis\cite{Standish12}, and when corrected, led to significant
complexity surpluses being generated for EcoLab and Tierra, though not
for WebWorld, another evolutionary ecology model similar to EcoLab.

In this paper I report on another (non-evolutionary) mechanical
technique for generating networks that does generate significant
amounts of complexity surplus.

\section{Generating networks from timeseries data}\label{network-generation}

Michael Small has been analysing timeseries data by generating
networks that represent the dynamics behind the timeseries, and then
applying network analysis techniques to tease out features of the
data. His most recent version of the technique\cite{Small13}, which I
shall describe below, can be applied to any timeseries, continuous or
discrete, and of any dimension. Of particular interest are timeseries
derived from chaotic dynamics, which generate particularly beautiful
filigree networks, that are prime candidates for high structural complexity.

\begin{figure}
\begin{center}
\includegraphics{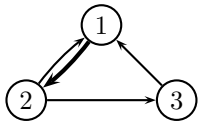}
\end{center}
\caption{Example network generated from the timeseries 1,2,3,1,2,1.}
\label{example-network}
\end{figure}

Consider first a discrete timeseries $(x_1,x_2,\ldots,x_n)$ where
$x_i\in X\subset\integers$. Each element of $X$ labels a node
of the target network, and links of the network are weighted by the
number of transitions $(x_i,x_{i+1})$ in the timeseries between the
labels of source and target  nodes of the network link. Concretely,
consider a simple timeseries 1,2,3,1,2,1. The generated network is
shown in figure \ref{example-network}, where the link $1\rightarrow 2$
has double the weight of the other links.

If the timeseries data is continuous rather than discrete, then a
means to convert real-valued data into integer labels is required. In
this paper, I use a simple coarse graining, where the number of cells
in each dimension of the timeseries is chosen to give a target network
size. For example, with three dimensional data, one can choose 10
cells along each dimension to give a target network size of 1000
nodes.

It should be pointed out that Small gives a more sophisticated
approach aimed at extracting the maximum dynamic information from the
timeseries, which involves choosing a window size $w$, and ranking the
values within the window to give a well defined sequence of
integers. For example, the sequence (0.5, 0.2, 0.35, 0.4, 0.3) will
for $w=3$ give the sequence of ranks: (3,1,2), (1,2,3), (2,3,1). The
generated networks will have $m(w)\le w!$ nodes. Finally, the
parameter $w$ is not arbitrary, but chosen to maximise the amount of
information captured. The curve $m(w)$ is roughly sigmoidal in shape,
and the optimum value $w_\mathrm{opt}$ occurs at the point of
inflection, where $\Delta m(w_\mathrm{opt})$ is maximised.

\section{Results}

\begin{table}
\begin{center}
\begin{tabular}{|l|r|r|r|r|r|r|}
\hline
Dataset & nodes & links &${\cal C}$ & $e^{\langle\ln{\cal C}_\mathrm{ER}\rangle}$ & ${\cal
  C}-e^{\langle\ln{\cal C}_\mathrm{ER}\rangle}$ & $\frac{|\ln{\cal
  C}-\langle\ln{\cal C}_\mathrm{ER}\rangle|}{\sigma_\mathrm{ER}}$ \\\hline
celegansneural & 297 & 2345 & 442.7 &251.6 &191.1 &29\\
PA1 & 100& 99 & 98.9& 85.4 & 13.5& 2.5\\
Lorenz & 8000 & 62 & 560.2 & 56.0&  504.2 & 58.3\\
H\'enon-Heiles & 10000&  31&  342.0&  57.3&  284.7&  55.6\\\hline
\end{tabular}
\end{center}
\caption{Node and link count, network complexity, average shuffled
complexity, surplus and the number of standard deviations of the
shuffled distribution that the surplus represents. Listed here are the
well-known {\em C. elegans} neural network, preferential attachment
with outdegree 1 (from \cite{Standish12}), and networks generated from
the Lorenz and H\'enon-Heiles systems.}
\label{dynSysResults}
\end{table}
  
The first study looked at applying Small's network generation
technique to the dynamics of a couple of well-known chaotic dynamical
systems --- the Lorenz system, given by equation (\ref{lorenz}) and
the H\'enon-Heiles system, given by equation (\ref{henonheiles}).
\begin{eqnarray}\label{lorenz}
\dot{x}&=&\sigma(y-x)\nonumber\\
\dot{y}&=&x(\rho-z)-y\\
\dot{z}&=&xy-\beta z\nonumber\\
\sigma=10, \rho&=&28, \beta=8/3\nonumber
\end{eqnarray}
\begin{eqnarray}\label{henonheiles}
\ddot{x}&=&-x-2 xy\nonumber\\
\ddot{y}&=&-y-(x^2-y^2)
\end{eqnarray}

The networks were generated according to the scheme described in
\S\ref{network-generation}. Shown in table \ref{dynSysResults} are the
node and link counts of the generated network, the complexity value
${\cal C}$ computed according to eq (\ref{complexity}), the average
complexity value of shuffled networks, the difference between the
complexity and the average shuffled complexity (ie surplus) and
finally the number of standard deviations of the shuffled distribution
that the surplus corresponds to. It is reported this way, as the $p$
value is far too ridiculously small to be comprehensible. For more
detailed discussion of the analysis, please refer to
\cite{Standish12}. Also shown, for comparison, are a couple of results
from that paper.

\begin{figure}
\begin{center}
\includegraphics{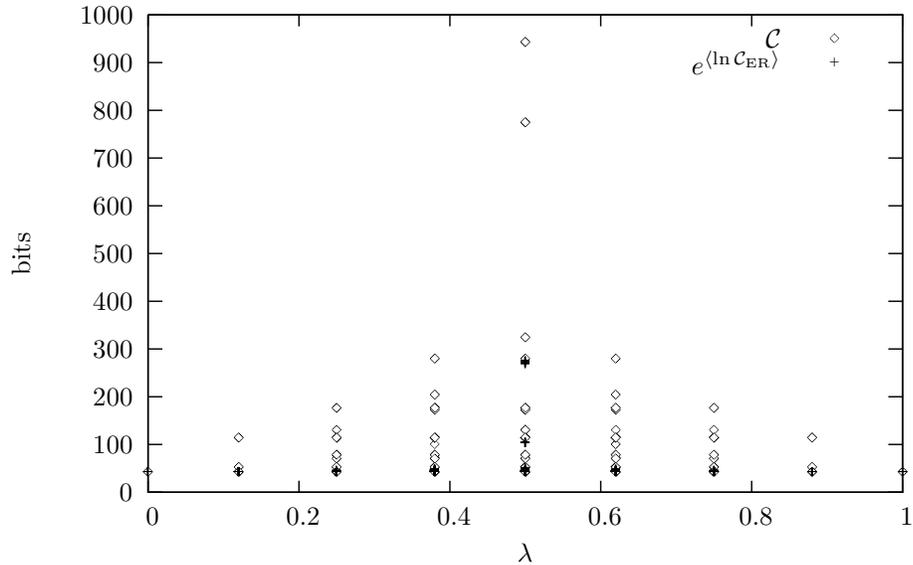}
\end{center}
\caption{Plot of ${\cal C}$ and averaged shuffled ${\cal C}$ for networks
generated from all 256 1D CAs with a 3 cell neighbourhood, plotted as
a function of Langton's $\lambda$.}
\label{1DCA}
\end{figure}

The next experiment performs the same analysis on timeseries generated
by 1D binary cellular automata with a neighbourhood of 3. The number
of cells chosen was 10, so the generated network will have 1024 nodes
($2^{10}$ possible states). The CAs were initialised to a random state,
and the first 1000 steps discarded to eliminate transient
effects. Figure \ref{1DCA} shows the complexity, and average shuffled
complexity plotted as a function of Langton's
$\lambda$\cite{Langton90}, for all 256 neighbourhood 3 binary CAs. What
is clear is that there a large spike in the complexity values (and in
the surplus) around $\lambda=0.5$, corresponding to chaotic rules.

Source code used for these experiments can be found as part of
\EcoLab{} version 5.D19, available from http://ecolab.sourceforge.net.

\section{Discussion}

Generating networks from chaotic dynamical systems proves to be a good
means of generating networks with complexity surplus. We can now
begin to study systems that can be tuned across a range of behaviour
from ordered to chaotic. A preliminary experiment showed that chaotic
behaviour is associated with the presence of large amounts of network
complexity. One might suspect that the complexity surplus is
associated with complex dynamics, or with Wolfram class 4 cellular
automata. The preliminary experiment reported here did not specifically support
that, as the peak occurred for chaotic (class 3) cellular automata,
however the class 4 regime tends to be a very small part of the
$\lambda$ spectrum, so I doubt that enumerating all 1D 3-neighbourhood
CA can resolve the issue. Instead, transition to chaos experiments,
such as Langton's original study with a 4-state,
5-neighbourhood CA\cite{Langton90} will probably be more suitable.

\section*{Acknowledgments}

I wish to thank Michael Small for helpful discussions on this topic.

\bibliographystyle{splncs03}
\bibliography{rus}

\end{document}